\documentclass[useAMS,usenatbib]{mn2e}
\usepackage{graphics}
\usepackage{latexsym}

\begin{document}
\title[Cataclysmic variables from the Cal\'an-Tololo Survey. I.]
{ Cataclysmic variables from the Cal\'an-Tololo-Survey -- I. Photometric 
periods}
\author[C. Tappert et al.]
{
C.~Tappert,$^1$\thanks{email: ctappert@astro.puc.cl}
T.~Augusteijn,$^2$ and 
J.~Maza$^3$ \\
$^1$Departamento de Astronom\'{\i}a y Astrof\'{\i}sica, Pontificia 
Universidad Cat\'olica, Casilla 306, Santiago 22, Chile\\
$^2$Nordic Optical Telescope, Apartado 474, E-38700 Santa Cruz de La Palma, 
Spain\\
$^3$Departamento de Astronom\'{\i}a, Universidad de Chile, Casilla 36-D,
Santiago, Chile
}

\maketitle

\begin{abstract}
 In a search for cataclysmic variables in the Cal\'an-Tololo Survey we have
 detected 21 systems, 16 of them previously unknown. In this paper we present 
 detailed time-series photometry for those six confirmed cataclysmic variables
 that show periodic variability in their light curves.
 Four of them turned out to be eclipsing systems, while the 
 remaining two show a modulation consisting of two humps. All 
 derived periods are below, or, in one case,
 just at the lower edge of the period gap. 
\end{abstract}
 

\section{Introduction}

Cataclysmic variables (CVs) are close interacting binaries where a white dwarf
primary accretes matter from a late-type main-sequence star. In absence of 
strong magnetic fields on the primary star, an accretion disc is formed, where
the matter transferred from the secondary star spirals inwards until it is
accreted by the white dwarf. 
For a comprehensive overview on CVs see \citet{warn95}.

The general evolution from the detached binary to the semi-detached state of a 
CV, and the subsequent long-term evolution to shorter orbital periods is
thought to be driven by orbital angular momentum losses due to gravitational 
radiation and magnetic braking. Theoretical models based on this scenario imply
that a CV should evolve down to an orbital period of $\sim$2 h in less than
10$^9$ yr.
The evolutionary lifetime of CVs near the shortest period 
($\sim$80 min) should become drastically longer, as the thermal timescale of
the secondary star becomes longer than the timescale of angular momentum loss 
by gravitational radiation. This behaviour should cause a ``spike'' in
the orbital period distribution at the minimum period, as systems
``pile up'' at this point \citep*{kolb93,steh+97,patt98}. The theoretical 
predictions for this 
minimum period range up to $\sim$70 min \citep{kolbbara99}. 
Population synthesis calculations predict a space density 
$\sim$10$^{-4}$ pc$^{-3}$, with 99\% of the CVs having periods shorter than 2 
h, and 70\% even having passed the turning point at the minimum period, with 
the now degenerate secondaries forcing an evolution towards longer orbital
periods \citep*{kolb93,steh+97,howe+97}.

However, there exists a strong mismatch between these predictions and
the actually observed sample. Observational estimates from large-area
surveys derive a space density roughly a factor 10 lower than the
predicted one \citep[][and references therein]{patt98}. The observed
period distribution of CVs shows only about half of the sources to
have an orbital period below the period gap \citep{rittkolb03}.
The minimum of the observed period distribution is at $\sim$78 min, and
there is no evidence of an enhancement in the number of sources close
to this period \citep[e.g.,][]{barkkolb03}. 

Alternative evolutionary models have been proposed 
\citep[e.g.,][]{kingsche02,sprutaam01}. More significantly,
the strength of the magnetic breaking, which is one of the fundamental
ingredients in the standard model, appears to be much weaker than assumed
\citep*{andr+03}. However, the discrepancies 
described above can (at least partly) be the result of the observational biases
affecting the sample of CVs, and this should be explored before discarding the 
standard (and in other respects quite successful) model.

In the standard model, the ``missing'' part of the population is
thought to be dominated by low mass-transfer rate, short period dwarf
novae, which are intrinsically very faint and show only very
infrequent outbursts. Especially the relatively bright magnitude
cut-off of large-scale surveys conducted hitherto 
\citep[Palomar-Green, Edinburgh-Cape Blue Object Survey:][respectively]{gree+86,stob+97} 
make them strongly biased against these systems 
\citep{patt84,shar+86}.

The low mass-transfer rate dwarf novae we are looking for are expected to
have a fairly blue continuum and strong hydrogen emission lines
\citep[see, e.g., Fig. 6 in][]{patt84}. This emphasises the
importance of using deep spectroscopic and photometric surveys for a
systematic search for CVs. First results from such surveys have become
available recently \citep*[e.g., the Hamburg Quasar Survey and the Sloan
Digital Sky Survey;][respectively]{gaen+02,szko+02,szko+03}.

Starting in 1996, we have conducted a survey of candidate CVs selected from the
Cal\'an-Tololo Survey \citep[CTS; see][and references therein]{maza+89}.  
The CTS has been designed to search for emission-line 
galaxies, quasars and galaxies with strong UV excess using objective prism 
plates. The spectra cover the wavelength range from $\sim$5300 {\AA} down to 
the atmospheric cut-off. The survey includes 5150 deg$^2$, i.e.\ $\frac{1}{8}$ 
of the whole sky, in the southern hemisphere with $|b| \geq 20^{\circ}$. The 
typical limiting magnitude of the plates is $B_J\sim$18.5. About half of the 
CTS plates ($\sim$2100 deg$^2$) had been examined in 1996, and those 
sources showing a blue continuum and hydrogen emission lines were selected as 
candidate CVs, yielding a sample of 59 objects. Follow-up observations 
identified 21 targets as definite CVs, including 16 previously unknown systems
of which 4 CVs have been independently discovered by other surveys in the
meantime \citep*{tapp+02}. 

In this paper we present detailed data for those CVs that show photometric 
modulations we believe to reflect the orbital periods. In future papers we will 
present time-resolved spectroscopy for the remaining CVs without previously
known periods, and discuss the complete sample.

\section{Observations and reductions}

\begin{table*}
\caption[]{The  photometric observations. $n_{\rm data}$ gives the number of 
data points, $t_{\rm exp}$ the individual exposure time, $\Delta t$ the time 
range covered by the observations, and $V_{\rm av}$ the average $V$ magnitude 
during the run.}
\label{phobs_tab}
\begin{tabular}{l l l l l l l l l l l}
\hline
CTCV & R.A.$_{\rm J2000}$ & DEC$_{\rm J2000}$ & date & HJD & telescope & filter
& $n_{\rm data}$ & $t_{\rm exp}$ [s] & $\Delta t$ [h] & $V_{\rm av}$ \\
\hline
J0549$-$4921 & 05:49:45.4 & $-$49:21:56 & 1996-10-05 & 2\,450\,362
& 0.9 m & $V$ & 168 & 20 & 2.55 & 13.75 \\
 & & & 1996-10-06 & 2\,450\,363 & 0.9 m & $V$ & 125 & 20 & 1.79 & 13.79 \\
 & & & 1996-10-07 & 2\,450\,364 & 0.9 m & $V$ & 65 & 20 & 0.93 & 14.52 \\
 & & & 1998-10-31 & 2\,451\,118 & 2.2 m & Gunn r & 184 & 20 & 3.16 & 16.86$^2$ \\
 & & & 1998-11-01 & 2\,451\,119 & 2.2 m & Gunn r & 224 & 20 & 3.71 & 16.86$^2$ \\
 & & & 1998-11-02 & 2\,451\,120 & 2.2 m & Gunn r & 246 & 20 & 4.24 & 16.92$^2$ \\
 & & & 1998-11-03 & 2\,451\,121 & 2.2 m & Gunn r & 242 & 20 & 4.25 & 17.10$^2$ \\
 & & & 1998-11-04 & 2\,451\,122 & 2.2 m & Gunn r & 244 & 20 & 4.07 & 17.09$^2$ \\
J1300$-$3052 & 13:00:29.1 & $-$30:52:57 & 1996-04-23 & 2\,450\,197
& 0.9 m & $V$ & 74 & 120 & 6.44 & 15.72$^1$ \\
 & & & 1996-04-24 & 2\,450\,198 & 0.9 m & $V$ & 154 & 60/120 & 4.88 & 15.53$^1$ \\
 & & & 1997-04-28 & 2\,450\,567 & 0.9 m & $V$ & 306 & 30/60 & 6.76 & 18.67$^1$ \\
J1928$-$5001 & 19:28:32.6 & $-$50:01:34 & 1996-07-08 & 2\,450\,273
& 0.9 m & $V$ & 55 & 240/80 & 3.09 & 17.79$^1$ \\
 & & & 1996-07-09 & 2\,450\,274 & 0.9 m & $V$ & 197 & 60 & 5.21 & 17.87$^1$ \\
 & & & 1996-07-30 & 2\,450\,295 & 2.2 m & $V$ & 47 & 15/120 & 3.12 & 17.80$^1$\\
J2005$-$2934 & 20:05:51.2 & $-$29:34:58 & 1996-10-05 & 2\,450\,362
& 0.9 m & $V$ & 169 & 30 & 2.90 & 15.91 \\
 & & & 1996-10-07 & 2\,450\,364 & 0.9 m & $V$ & 164 & 30 & 2.81 & 15.92 \\
J2315$-$3048 & 23:15:31.9 & $-$30:48:46 & 1996-10-06 & 2\,450\,363
& 0.9 m & $V$ & 124 & 60 & 3.14 & 16.83$^1$ \\
J2354$-$4700 & 23:54:20.4 & $-$47:00:20 & 1996-10-03 & 2\,450\,360
& 0.9 m & $V$ & 24 & 240 & 1.78 & 18.87$^1$ \\
 & & & 1996-10-04 & 2\,450\,361 & 0.9 m & $V$ & 66 & 180 & 3.83 & 19.02$^1$ \\
\hline
\multicolumn{10}{l}{(1) out-of-eclipse value, (2) Gunn $r$ magnitude} \\
\end{tabular}
\end{table*}

\begin{table*}
\caption[]{The spectroscopic observations. Columns include the exposure time 
$t_{\rm exp}$, the wavelength range $\Delta \lambda$, the resolution 
$\Delta \lambda_{\rm FWHM}$, and the $V$ magnitude of the object at the time of
the spectroscopic observations. The latter was determined from the acquisition 
frames.}
\label{spobs_tab}
\begin{tabular}{l l l l l l l l l}
\hline
CTCV & date & tel. / instr. & grism & slit [\arcsec] & $t_{\rm exp}$ [s] 
& $\Delta \lambda$ [{\AA}] & $\Delta \lambda_{\rm FWHM}$ [{\AA}] & V [mag] \\
\hline
J0549$-$4921 & 1996-10-09 & 1.54 m / DFOSC & 11 & 1.5 & 5400 
& 4340 -- 9750 & 16 & 16.351(15) \\
J1300$-$3052 & 1996-07-31 & 2.2 m / EFOSC2 & 1 & 1.0 & 3600 
& 3800 -- 10500 & 30 & 18.730(33) \\
J1928$-$5001 & 1996-07-30 & 2.2 m / EFOSC2 & 1 & 1.0 & 3600
& 3800 -- 10500 & 30 & 17.834(12) \\
J2005$-$2934 & 1996-07-30 & 2.2 m / EFOSC2 & 1 & 1.0 & 600
& 3800 -- 10500 & 30 & 15.349(06) \\
J2315$-$3048 & 1996-07-30 & 2.2 m / EFOSC2 & 1 & 1.0 & 1200
& 3800 -- 10500 & 30 & 16.960(21) \\
J2354$-$4700 & 1996-07-29 & 2.2 m / EFOSC2 & 1 & 1.0 & 2$\times$900
& 3800 -- 10500 & 30 & 19.70$^1$\\
\hline\noalign{\smallskip}
\multicolumn{7}{l}{(1) spectrophotometric magnitude, not corrected for slit
losses} \\
\end{tabular}
\end{table*}

Follow-up observations for all objects in the original sample were conducted
at ESO telescopes in the form of optical spectroscopy, calibrated and 
time-series photometry. The dates and instruments used for the observations 
presented here are summarised in Tables \ref{phobs_tab} and \ref{spobs_tab}.

The photometric data were obtained, with one exception, at the 0.9 m Dutch
telescope, which was equipped with a 512$\times$512 pixel CCD. 
One observing run for CTCV J0549$-$4921 was conducted on the ESO/MPI 2.2 m 
using a test camera equipped with a Gunn r filter.
Reduction was performed using IRAF routines for
BIAS and flatfield correction. Aperture photometry was obtained with IRAF's
daophot package \citep{stet92}, and the differential light curves
were computed with respect to one or more suitable comparison stars that were
checked for constancy.

Calibrated magnitudes for the systems were determined by single B, V, R
exposures and subsequent comparison with suitable (i.e., non-variable) 
stars within the target's CCD frame, which in return were calibrated with
respect to \citet{land92} standards. The colour information derived from
the calibrated photometry will be analysed in the concluding paper.

The spectroscopic data were taken at the Danish 1.54 m and the ESO/MPI 2.2 m
telescopes. After bias and flatfield correction the spectra were calibrated in 
wavelength using Helium-Neon (1.54 m telescope) and Helium-Argon (2.2 m
telescope) lamps. Flux calibration was performed with respect to standard stars
observed in the same night. 
The flux calibrated spectra were folded with \citet{bess90} filter curves to 
obtain photometric magnitudes. By comparison with the values derived from
the acquisition frames (typically an exposure in the V or R filter) we 
determined the slit losses. Extreme values ranged from 0.2 mag 
(CTCV J1300$-$3052) to 1.1 mag (CTCV J0549$-$4921) with most values around 
0.4 mag. Folding the spectra with more than one filter furthermore yielded 
photometric colours. Comparison with BVR photometry taken of the target in a 
similar brightness state showed an agreement in colour within 0.05 mag. For 
CTCV J2354$-$4700, no acquisition frame was available, hence the accuracy of 
the flux calibration could not be tested in this case.

\section{Results}

\begin{figure*}
\resizebox{18cm}{!}{\includegraphics{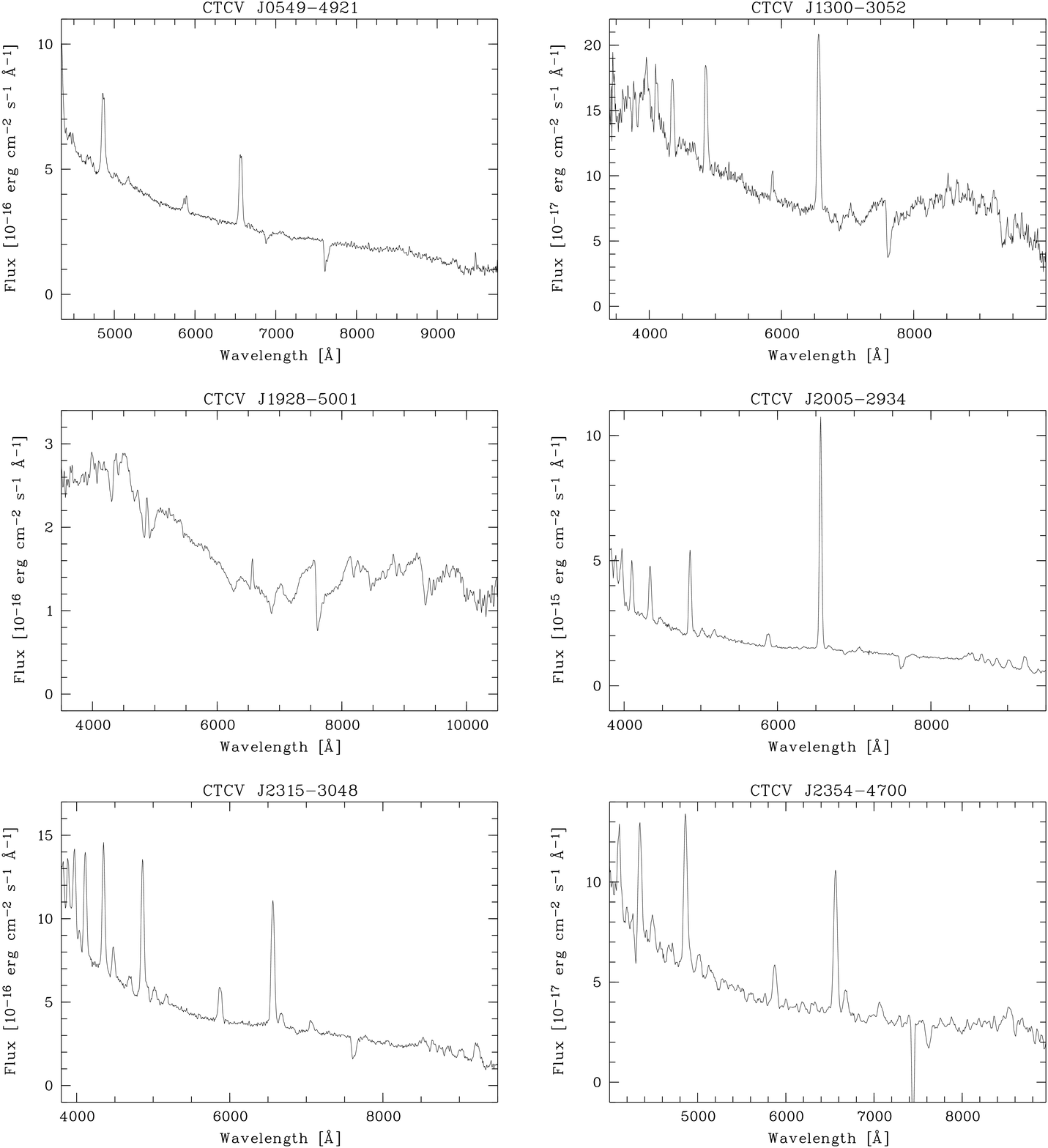}}
\caption[]{Spectra of the 6 CTCV sources as indicated. The absorption feature
at $\lambda$7440 for CTCV J2354$-$4700 is an artifact.}
\label{sp_fig}
\end{figure*}

\begin{table*}
\caption[]{Equivalent widths (in {\AA}) of the most prominent emission lines.
A blank field indicates that this part of the spectrum was not covered
with our observations, a `$-$' that the corresponding line was not detected.}
\label{eqw_tab}
\begin{tabular}{l l l l l l l l l l l l l l l}
\hline
CTCV & \multicolumn{4}{c|}{Balmer} & \multicolumn{5}{c|}{He\,{\sc I}}
& He\,{\sc II} & Fe\,{\sc II} \\
 & 4102 & 4340 & 4861 & \multicolumn{1}{l|}{6563}
& 4471 & 5016 & 5876 & 6678 & \multicolumn{1}{l|}{7065} & 4686 & 5173\\
\hline
J0549$-$4921 &    &    & 35 & 52  & 6  & 4  & 11 & 5  & -- & 8  & 6  \\
J1300$-$3052 & 15 & 24 & 47 & 110 & -- & -- & 12 & -- & -- & -- & -- \\
J1928$-$5001 & -- & -- & 13 & 11  & -- & -- & -- & -- & -- & -- & -- \\
J2005$-$2934 & 25 & 33 & 66 & 219 & 6  & 8  & 18 & 5  & 9  & -- & 8  \\
J2315$-$3048 & 31 & 44 & 45 & 102 & 14 & 11 & 32 & 11 & 11 & 9  & 6  \\
J2354$-$4700 & 13 & 41 & 58 & 104 & 22 & 21 & 40 & 21 & 18 & -- & -- \\
\hline
\end{tabular}
\end{table*}

For the majority of the systems we have measured the arrival times of humps or
eclipses by fitting a polynomic function of second order to the periodic
features. Since the uncertainties of the arrival times resulting from the 
parabolic fit are usually smaller than the real variation (e.g., due to changes
in the profile of the feature caused by flickering), we have assumed identical 
uncertainties for all arrival times for a given source and scaled them to give
a reduced $\chi^2 = 1.0$ for the fit to the arrival times to determine the
ephemeris. For these sources with a well defined feature in their light curve, 
the cycle count between the times of arrival of the feature were unambiguous 
and the period could be derived directly.

In the two cases where this approach was not possible, we have computed
periodograms based on the analysis-of-variance (AOV) algorithm 
\citep{schw89} as implemented in MIDAS. A Monte Carlo 
simulation was applied in order to test the significance of alias peaks and to 
obtain an error estimation for the periods \citep{tappbian03,menntapp01}.

\subsection{CTCV J0549$-$4921\label{j0549_sec}}

The spectroscopic appearance of CTCV J0549$-$4921 is that of a typical dwarf
nova, with moderately strong H and He\,{\sc I} emission (Table \ref{eqw_tab}). 
No signatures of the secondary star are detected (Fig.\ \ref{sp_fig}). 

\begin{figure}
\resizebox{8cm}{!}{\includegraphics{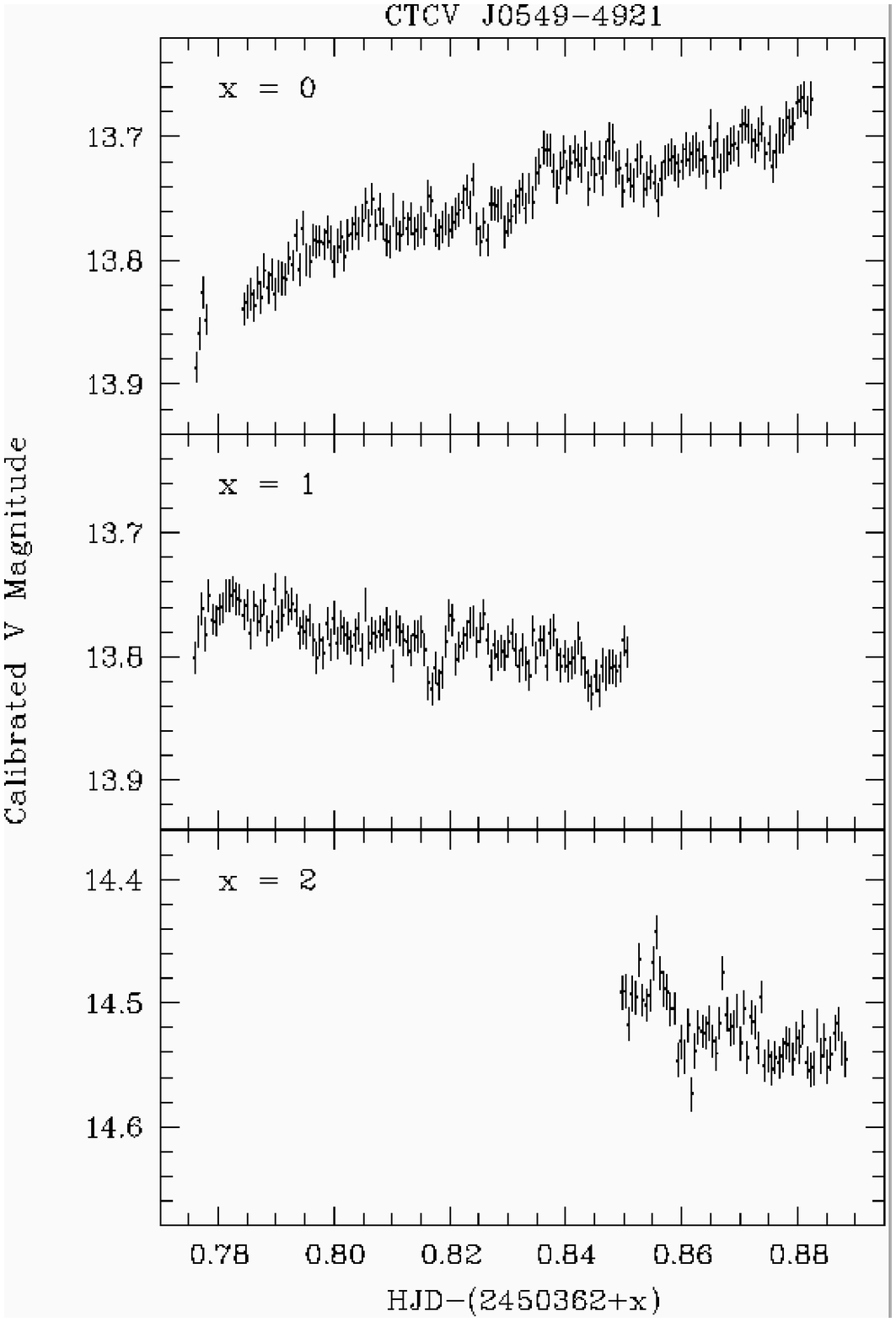}}
\caption[]{Light curves for the 1996 data of CTCV J0549$-$4921. Note that the
relative magnitude range is the same for all plots.}
\label{j0549lc96_fig}
\end{figure}

\begin{figure}
\resizebox{8.2cm}{!}{\includegraphics{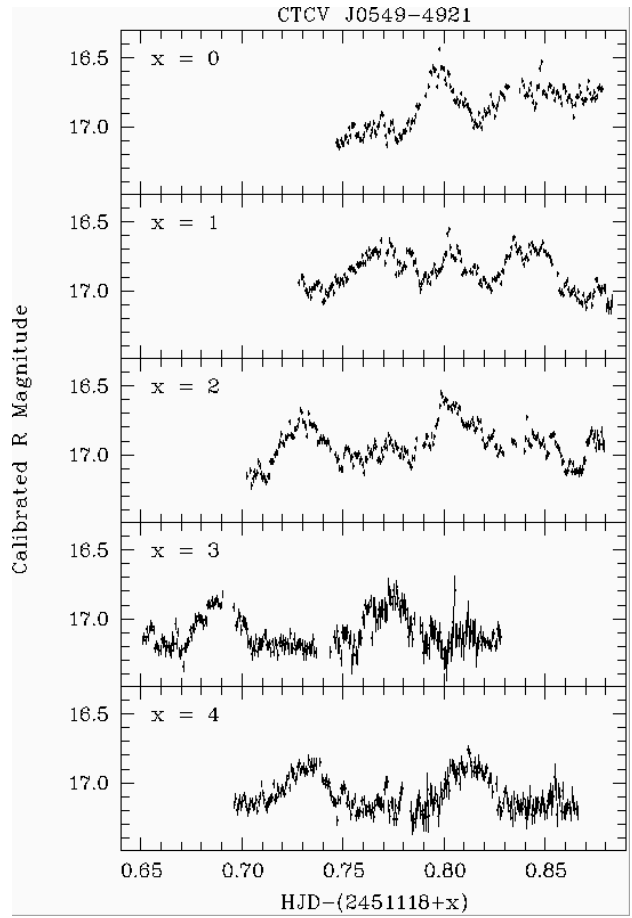}}
\caption[]{Light curves for the 1998 data of CTCV J0549$-$4921.}
\label{j0549lc98_fig}
\end{figure}

Time-series photometry of the system was taken on two occasions. During the 
first run in 1996, it proved to be in outburst, without showing any obvious 
periodic features (Fig.\ \ref{j0549lc96_fig}). The second run, in 1998, 
caught the CV in a $\sim$2.2 mag fainter state \citep[we have no information on
the colour $V-R$, but typical values for CVs are 0.1--0.5 mag;][]{eche84}.
The light curves (Fig.\ \ref{j0549lc98_fig}) over 5 nights show little 
variation of the average value, and we therefore assume that the system was in,
or close to, quiescence. The most prominent feature in these light curves is a 
periodic hump with an amplitude of $\sim$0.5 mag. A 
second, smaller and more distorted, hump is also present. 

\begin{figure}
\rotatebox{270}{\resizebox{6cm}{!}{\includegraphics{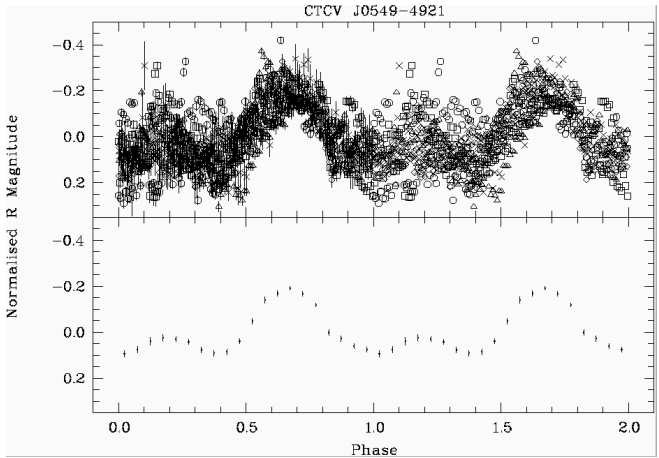}}}
\caption[]{Light curve for CTCV J0549$-$4921 folded on $P$ = 0.080\,239 d. The
zero point of the phase has been set to the first data point. The symbols in 
the upper plot refer to the individual data sets from 1998-10-31 ({\large 
$\circ$}), 11-01 ($\Box$), 11-02 ($\triangle$), 11-03 ($\times$), and 11-04 
($\Diamond$). The error bars from phase 0 to 1 correspond to the photometric 
error as in Fig.\ \ref{j0549lc98_fig}. The lower plot shows the average light 
curve. The bin size is 0.05 phases.} 
\label{j0549avlc_fig}
\end{figure}

We have measured the arrival times of the maximum of the primary hump by
fitting a second order polynomial function to the hump. The results are
given in Table \ref{j0549hump_tab}. The linear regression yields the ephemeris
\begin{equation}
T_0 = {\rm HJD}~2\,451\,118.8019(22) + 0.080\,218(70)~E,
\end{equation}
with $E$ giving the cycle number.
Fig.\ \ref{j0549avlc_fig} presents the 1998 light curve folded on this period.

Light curves containing variable double humps represent a phenomenon commonly 
seen in low mass-transfer rate dwarf novae 
\citep[e.g.,][for WZ Sge and BW Scl, respectively]{patt80,auguwiso97}. Within 
our sample we find corresponding 
features also for CTCV J2005$-$2934, CTCV J2315$-$3048, and CTCV J2354$-$4700.
In the latter two systems, the additional presence of an eclipse proves
the orbital nature of the double hump. The usual explanation is that the 
humps correspond to the bright spot that is caused by the impact of the gas 
stream from the secondary star on the accretion disc, and which is seen at 
inferior (primary hump) and superior (secondary hump) conjunction. Within this 
picture, our (arbitrary) phase of the maximum of the primary hump in CTCV 
J0549$-$4921 should translate to an orbital phase of $\sim$0.8 with respect
to superior conjunction of the white dwarf.

\begin{table}
\caption[]{Arrival times of the primary hump for CTCV J0549$-$4921. The
uncertainties are those of the parabolic fit. The estimated uncertainty
scaled with the reduced $\chi^2$ for all arrival times is $3.4\times10^{-3}$ d 
(see text).}
\label{j0549hump_tab}
\begin{tabular}{r l} 
\hline
cycle & HJD \\
\hline
0  &  2\,451\,118.7983(05) \\
12 &  2\,451\,119.7701(13) \\
13 &  2\,451\,119.8417(08) \\
24 &  2\,451\,120.7306(09) \\
25 &  2\,451\,120.8063(07) \\
36 &  2\,451\,121.6889(12) \\
37 &  2\,451\,121.7729(19) \\
49 &  2\,451\,122.7310(14) \\
50 &  2\,451\,122.8122(18) \\
\hline
\end{tabular}
\end{table}

\subsection{CTCV J1300$-$3052}

\begin{figure}
\resizebox{8cm}{!}{\includegraphics{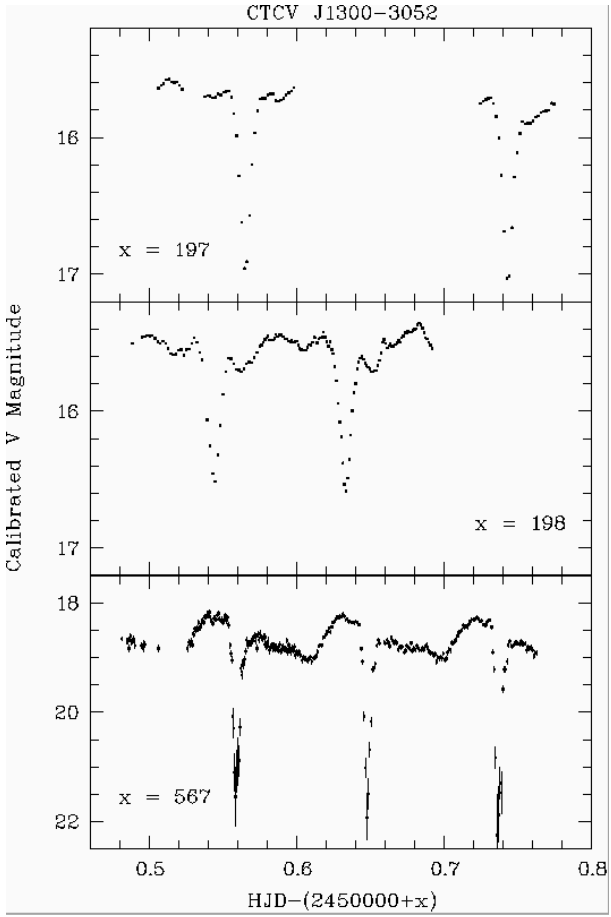}}
\caption[]{Light curves for CTCV J1300$-$3052 from 1996-04-23 (top), 1996-04-24
(middle), and 1997-04-28 (bottom). Note that the y-axis of the bottom plot
has a different scale.}
\label{j1300lc_fig}
\end{figure}

\begin{table}
\caption[]{Eclipse timings for  CTCV J1300$-$3052. Due to the cycle ambiguity
between both data sets, each refers to its individual zero point. The
estimated error scaled by the reduced $\chi^2$ for the 1996 data amounts to
$3.4\times10^{-4}$ d. For the 1997 data, the errors resulting from the
parabolic fit already gave reasonable residuals.} 
\label{j1300ecl_tab}
\begin{tabular}{r l}
\hline
cycle & HJD \\
\hline
1996: & \\
0  & 2\,450\,197.565\,12(05) \\
2  & 2\,450\,197.743\,57(06) \\
11 & 2\,450\,198.543\,86(08) \\
12 & 2\,450\,198.633\,29(08) \\
\hline
1997: & \\
0  & 2\,450\,567.559\,19(35) \\
1  & 2\,450\,567.648\,05(22) \\
2  & 2\,450\,567.736\,90(34) \\
\hline
\end{tabular}
\end{table}

\begin{figure}
\rotatebox{270}{\resizebox{6cm}{!}{\includegraphics{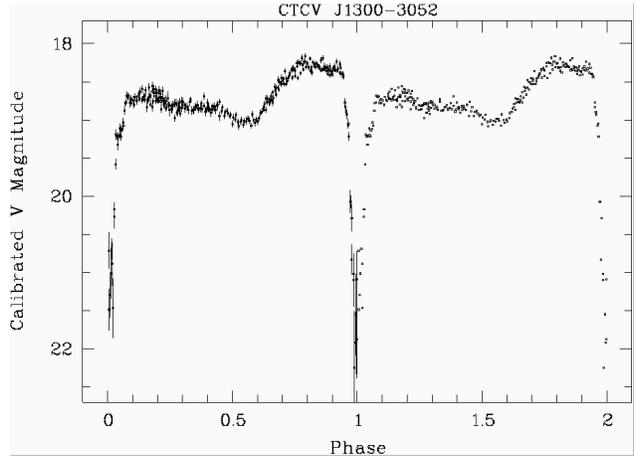}}}
\caption[]{The 1997 light curve for CTCV J1300$-$3052 folded on the ephemeris
given in Eq.\ \ref{j1300eph2_eq}.} 
\label{j1300phi_fig}
\end{figure}

\begin{figure}
\rotatebox{270}{\resizebox{6cm}{!}{\includegraphics{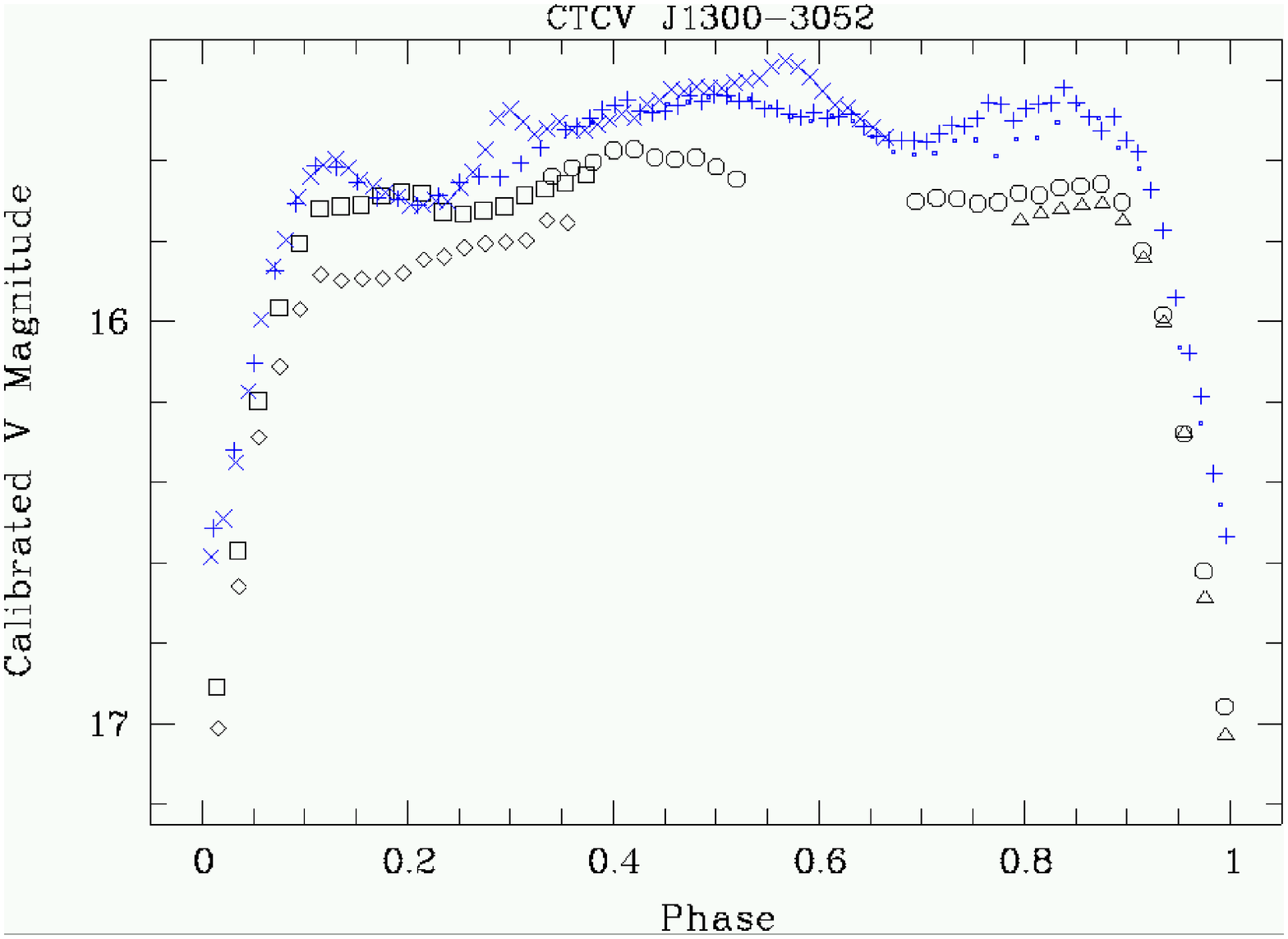}}}
\caption[]{Phase folded outburst data CTCV J1300$-$3052 with respect to the 
ephemeris given in Eq.\ \ref{j1300eph1_eq}. Different symbols indicate 
different orbital cycles: $\circ$, $\Box$, $\triangle$, $\diamond$, for cycles
$-1$, 0, 1, 2 (1996-04-23), and dot, $+$, $\times$, for cycles 10, 11, 12
(1996-04-24), respectively.}
\label{j1300out_fig}
\end{figure}

This object shows a composite spectrum with features from all three 
components: the typical strong emission lines from the accretion disc, which in
the blue part of the spectrum are embedded in broad, shallow absorption troughs
from the white dwarf primary, and the red continuum and absorption bands from
the secondary star (Fig.\ \ref{sp_fig}). A rough comparison with
standard star catalogues \citep*{jaco+84,silvcorn92} indicates an early M
type for the latter.
 
We have observed the system photometrically in two different years, which 
proved it to be in a state of increased brightness during the first observing 
run in 1996. The light curves of both runs are given in 
Fig.\ \ref{j1300lc_fig}. It shows the system with a $\sim$3 mag deep eclipse in
quiescence. In the system's high state the eclipse depth is diminished to 
$\sim$1.1 mag.

We have fitted second order polynomial functions to the eclipse data in order
to determine the times of mid-eclipse. The results are given in Table 
\ref{j1300ecl_tab}. Unfortunately, a period analysis could not be performed
on the combined data set, since the time span between the two observing 
runs proved too large for an unambiguous cycle count. Linear fits to the
mid-eclipse timings for the individual runs yielded the ephemerises
\begin{equation}
T_0 = {\rm HJD}~2\,450\,197.565\,33(26) + 0.088\,981(32)~E,
\label{j1300eph1_eq}
\end{equation}
for the 1996 data, and
\begin{equation}
T_0 = {\rm HJD}~2\,450\,567.559\,19(29) + 0.088\,85(24)~E,
\label{j1300eph2_eq}
\end{equation}
for the 1997 data, with $E$ giving the cycle number. 
Within the errors, both periods are consistent with each other.
This period places the system close to the lower edge of the CV period gap 
\citep[e.g.\ Fig.\ 13 in][]{tappbian03}.

The quiescence data folded on its ephemeris shows a `classical' shape of a
broad hump prior to eclipse (Fig.\ \ref{j1300phi_fig}), very similar e.g.\ to 
that of U Gem \citep{krze65}. The outburst light
curve shows large, apparently irregular, variations from cycle to cycle 
(Fig.\ \ref{j1300out_fig}). 
A noteworthy feature of these data is the smaller depth and the
larger width of the eclipse with respect to the quiescence data ($\sim$ 1 mag
and 30 min for the full eclipse width compared to $\sim$ 3 mag and 19 min), as 
a consequence of increased disc luminosity and radius. 

A more detailed study of the eclipse light curve and additional follow-up 
spectroscopic observations will be published elsewhere (Augusteijn et al., in 
preparation).

\subsection{CTCV J1928$-$5001}

\begin{figure}
\rotatebox{270}{\resizebox{3.6cm}{!}{\includegraphics{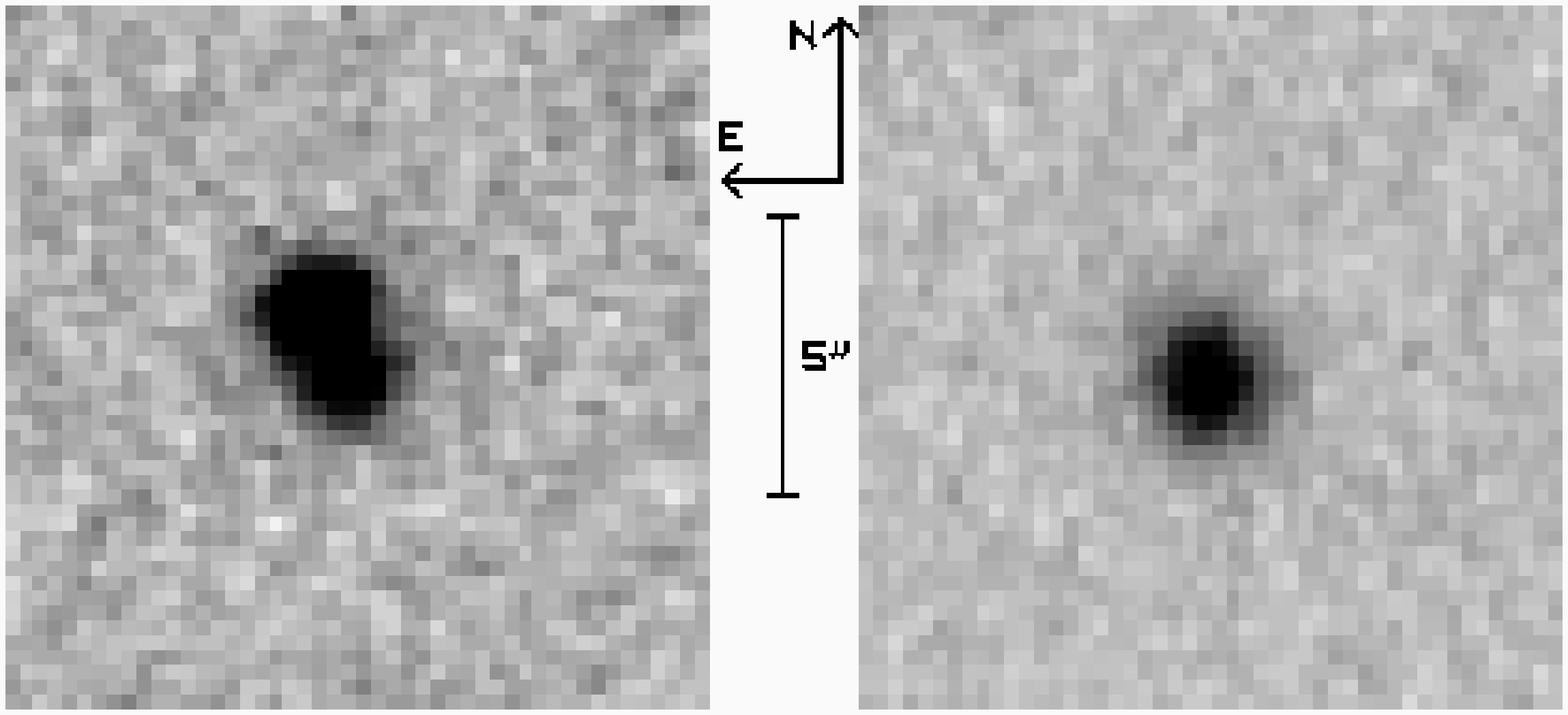}}}
\caption[]{Out-of-eclipse (left) and mid-eclipse (right) $V$ image of
CTCV J1928$-$5001 from 1996-07-30. 
}
\label{j1928ecl_fig}
\end{figure}

\begin{figure}
\rotatebox{270}{\resizebox{6cm}{!}{\includegraphics{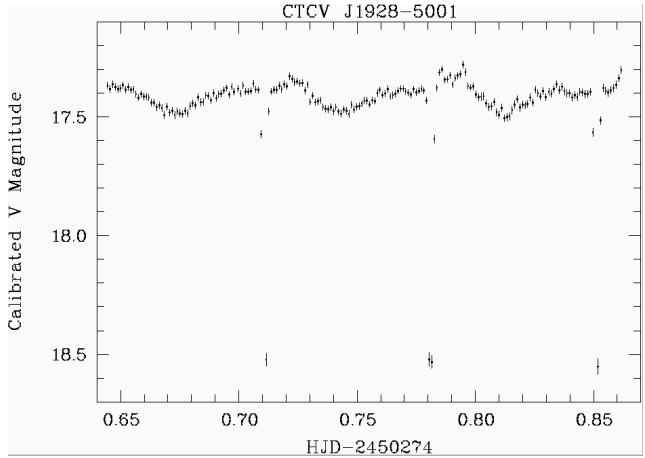}}}
\caption[]{Light curve of CTCV J1928$-$5001 from 1996-07-09. The magnitudes
represent the combined data from the CV and its close visual companion
($V$ = 18.5)} 
\label{j1928lc_fig}
\end{figure}

\begin{table}
\caption[]{Eclipse timings for  CTCV J1928$-$5001 as measured with a polynomial
fit of second order. The estimated uncertainty scaled by the reduced $\chi^2$
is $1.8\times10^{-4}$ d.}
\label{j1928ecl_tab}
\begin{tabular}{r l}
\hline
cycle & HJD \\
\hline
$-$15 & 2\,450\,273.728\,37(14) \\
$-$14 & 2\,450\,273.798\,92(08) \\
$-$1  & 2\,450\,274.711\,02(15) \\
0     & 2\,450\,274.781\,39(08) \\
1     & 2\,450\,274.851\,29(16) \\
297   & 2\,450\,295.624\,47(09) \\
298   & 2\,450\,295.694\,26(06) \\
\hline
\end{tabular}
\end{table}

\begin{figure}
\resizebox{8cm}{!}{\includegraphics{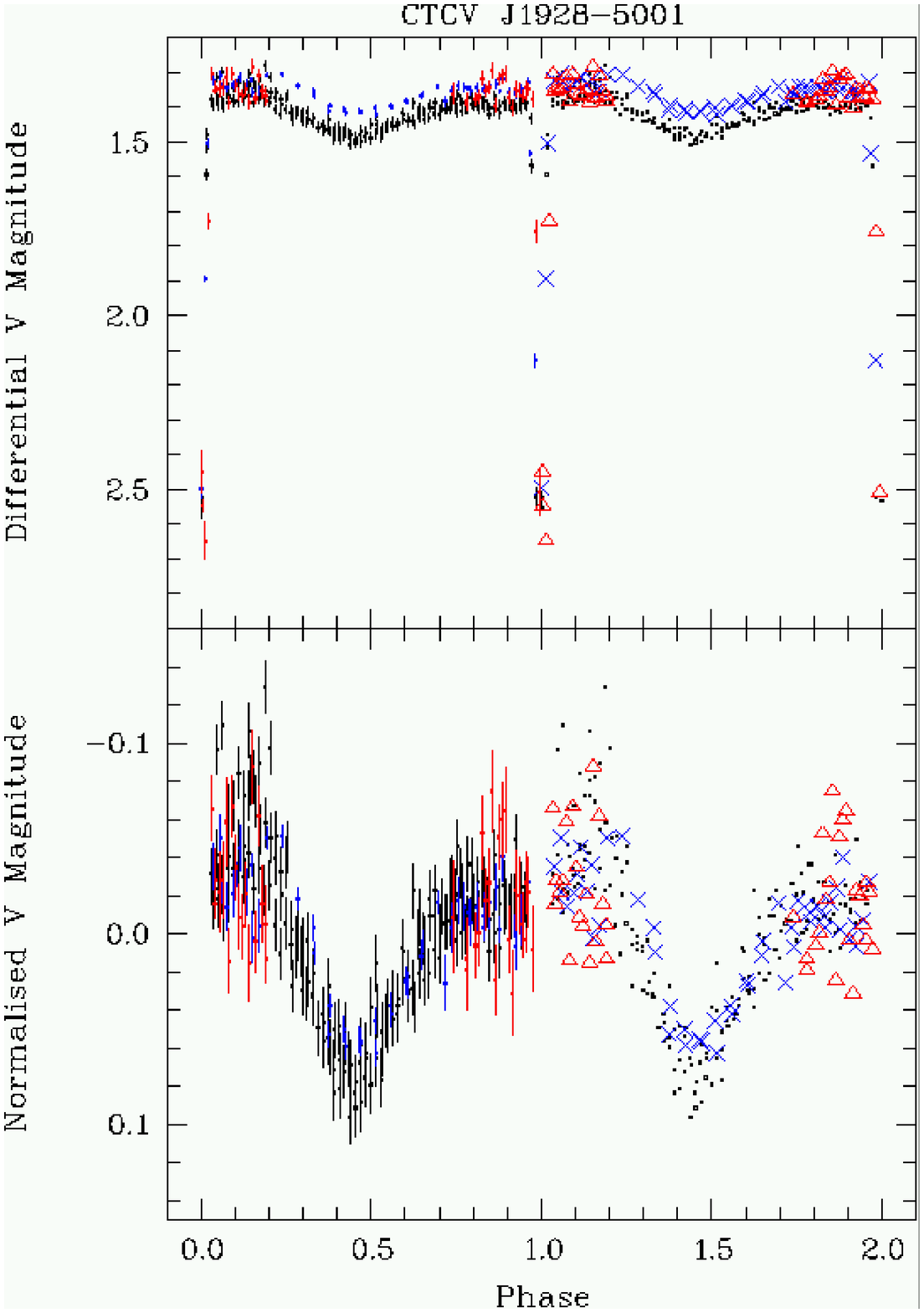}}
\caption[]{All three light curves of CTCV J1928$-$5001 folded on the ephemeris 
given in Eq.\ \ref{j1928eph_eq}. The upper plots shows the differential
magnitudes (the CV plus companion) to give an idea of the brightness variations 
of the system in its low state. The lower plot  represents a close-up of the 
out-of-eclipse data.The latter light curves have been normalised individually 
by subtracting the corresponding out-of-eclipse average magnitude. In both 
plots the symbols $\times$ and $\triangle$ from phases 1 to 2 refer to the 
1996-07-08 and 1996-07-30 data, respectively, while the dots represent the 
1996-07-09 data. The error bars from phases 0 to 1 correspond to the individual
photometric errors.} 
\label{j1928phi_fig}
\end{figure}

The spectrum of this object (Fig.\ \ref{sp_fig}) also shows contributions from
multiple components. In the red part we see the signatures of the secondary
star whose spectral type appears to be slightly later than M5V \citep{jaco+84}.
The blue part is dominated by the Balmer absorption lines from the white dwarf 
primary. The source was identified as a Polar (or AM Her) type CV by follow-up 
spectroscopic and polarimetric observations which will be published elsewhere 
(Potter et al., in preparation).

The object is a close visual binary with a separation $\approx$ 2\farcs4.
Additionally, the CV is a deeply eclipsing system (Figs.\ \ref{j1928ecl_fig}, 
\ref{j1928lc_fig}). Mid-eclipse photometry of the companion gave $V$ = 18.52(1)
mag, yielding out-of-eclipse values for the CV of $V$ = 17.7--18.0 mag. In
order to estimate the mid-eclipse magnitude of the target, we subtracted the
average PSF from the companion on an eclipse frame. Aperture photometry
on the residual yielded $V$ = 21.05(30). Note, however, that the error is a
purely photometric one, and does not take into account uncertainties in the
PSF fitting or in the time deviation from true mid-eclipse, especially since
the exact shape of the eclipse light curve is unknown. Therefore, 
deep mid-eclipse photometry will be needed to accurately determine the eclipse 
depth.
 
A linear fit to the timings of mid-eclipse (Table \ref{j1928ecl_tab}) yielded
an ephemeris
\begin{equation}
T_0 = {\rm HJD}~2\,450\,274.781\,24(08) + 0.070\,179(01)~E,
\label{j1928eph_eq}
\end{equation}
with the error of the zero cycle referring to the error of the linear
regression.
Fig.\ \ref{j1928phi_fig} shows the corresponding light curve.

\subsection{CTCV J2005$-$2934}

\begin{figure}
\rotatebox{270}{\resizebox{6cm}{!}{\includegraphics{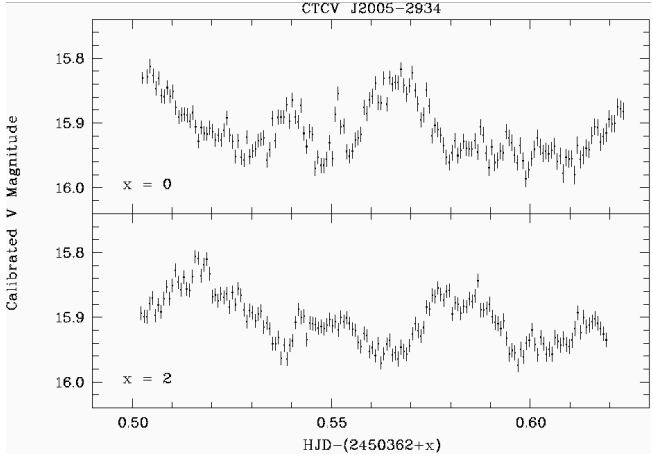}}}
\caption[]{Light curves for CTCV J2005$-$2934 from 1996-10-05 (top) and
1996-10-07 (bottom).} 
\label{j2005lc_fig}
\end{figure}

\begin{table}
\caption[]{Frequencies $f$, corresponding periods $P$, and the discriminatory
power $p_d$ from the Monte Carlo simulation on the CTCV J2005$-$2934 data.}
\label{j2005freq_tab}
\begin{tabular}{l l l l}
\hline
$f$ [cyc d$^{-1}$] & $P$ [d] & $p_d$ & ID\\
\hline
14.901(15) & 0.067\,111(69) & 0.000 & $P_1$ \\
15.398(05) & 0.064\,943(21) & 0.091 & $P_2$ \\
15.902(09) & 0.062\,887(37) & 0.637 & $P_3$ \\
16.401(04) & 0.060\,972(14) & 0.272 & $P_4$ \\
16.892(07) & 0.059\,200(25) & 0.000 & $P_5$ \\
\hline
\end{tabular}
\end{table}

\begin{figure}
\rotatebox{270}{\resizebox{6cm}{!}{\includegraphics{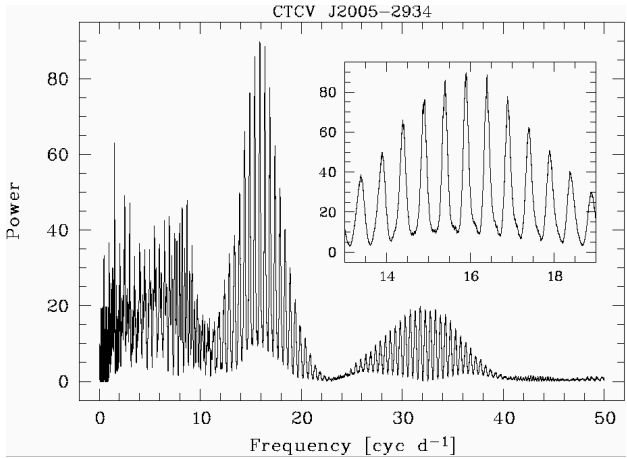}}}
\caption[]{AOV periodogram for the CTCV J2005$-$2934 data. The inlet shows a 
close-up of the main peaks.} 
\label{j2005pg_fig}
\end{figure}

\begin{figure}
\rotatebox{270}{\resizebox{6cm}{!}{\includegraphics{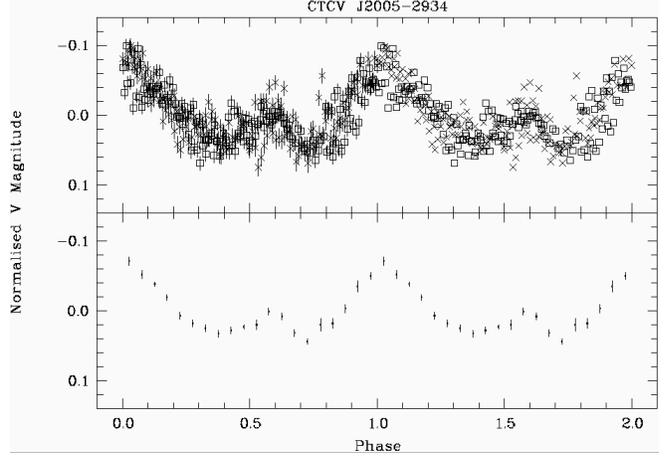}}}
\caption[]{Normalised light curve for CTCV J2005$-$2934 folded on $P$ = 
0.062\,887 d. The zero point of the phase has been set to the first data 
point. The symbols in the upper plot refer to the individual data sets from 
1996-10-05 ($\times$) and 1996-10-07 ($\Box$). The error bars from phase 0 to 1
correspond to the photometric error as in Fig.\ \ref{j2005lc_fig}. The lower 
plot shows the average light curve. The bin size is 0.05 phases.} 
\label{j2005phi_fig}
\end{figure}

Although the spectrum of this system displays all basic ingredients of a 
typical dwarf nova, it is peculiar in that its H$\alpha$ line is unusually 
strong, especially in comparison to the other Balmer lines (Fig.\ \ref{sp_fig},
Table \ref{eqw_tab}).

The photometric data was taken in two nights, with each data set spanning a 
very similar time range. The light curves are presented in Fig.\ 
\ref{j2005lc_fig}. Other than in the CTCV J0549$-$4921 data, the time range and
the number of covered humps are insufficient to use the arrival times of the
principal hump to derive the period. Additionally the hump appears less
constant in shape than the one of CTCV J0549$-$4921. After subtracting the 
nightly averages, we therefore applied the AOV algorithm to the combined data 
set. The corresponding periodogram shows a number of sharp peaks centred on 
$f \sim$16 cyc d$^{-1}$. For our Monte Carlo simulation we compared the five 
highest peaks. The corresponding frequencies and the resulting discriminatory 
power for a specific frequency over the other four are listed in Table 
\ref{j2005freq_tab}. Thus, $P_3$ emerges as the most probable period, while we 
can certainly discard $P_1$ and $P_5$. However, especially $P_4$ still 
represents an important alias, and additional observations will be necessary to
properly assign the orbital period.

In Fig.\ \ref{j2005phi_fig} we present individual light curves and the average
one folded on the most probable period $P_3$. Phase-folding on $P_2$ or $P_4$
yields very similar scatter and basically identical phasing of the main 
features. The light curve is in principle similar to that of CTCV J0549$-$4921 
in that it consists of two humps of different size. Their amplitudes, however, 
are only half as strong as those in CTCV J0549$-$4921, which could mean that 
CTCV J2005$-$2934 is seen at a somewhat lower inclination.
Using the same interpretation for the nature of these humps, orbital phase
$\sim$0.8 should correspond to our arbitrary phase 0.03 for the maximum of the 
primary hump.

\subsection{CTCV J2315$-$3048}

\begin{figure}
\rotatebox{270}{\resizebox{6cm}{!}{\includegraphics{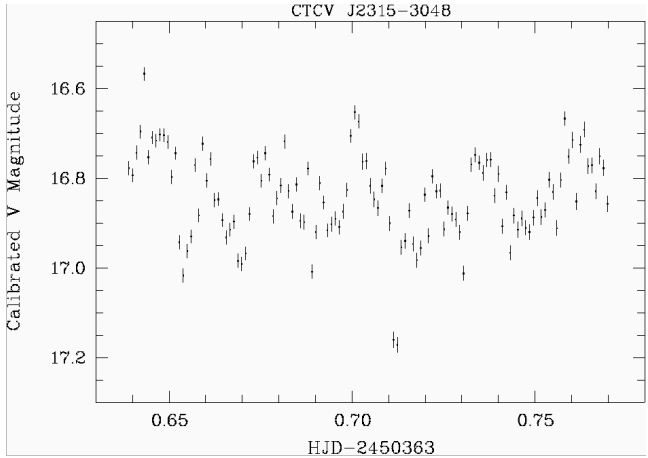}}}
\caption[]{Light curve for CTCV J2315$-$3048 from 1996-10-06.} 
\label{j2315lc_fig}
\end{figure}

\begin{figure}
\rotatebox{270}{\resizebox{6cm}{!}{\includegraphics{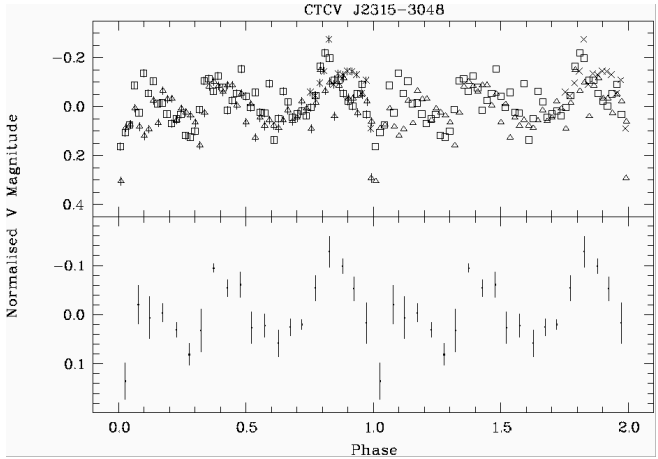}}}
\caption[]{Light curve for CTCV J2315$-$3048 folded on the ephemeris given
in Eq.\ \ref{j2315eph_eq}. The data have been normalised by subtracting a
polynomial of second order. For the fit the eclipse data and the first hump
were excluded. The data set in the upper plot has been divided into three 
parts, each corresponding to a different orbital cycle ($-$2 = $\times$, $-$1 =
$\Box$, 0 = $\triangle$). The error bars from phase 0 to 1 correspond to the 
photometric error as in Fig.\ \ref{j2315lc_fig}. The lower plot shows the 
average light curve with a bin size of 0.05 phases. The y-scale has been set in
such a way to enable direct comparison to Fig.\ 3 of \citet{chen+01}.} 
\label{j2315phi_fig}
\end{figure}

This object shows a very typical spectrum of a dwarf nova in quiescence with
strong emission lines of the H and He series (Fig.\ \ref{sp_fig}, Table
\ref{eqw_tab}). It has been independently discovered as the ROSAT source
1RXS J231532.3$-$304855 \citep{schw+00} and in the
Edinburgh-Cape Blue Object Survey as EC 23128$-$3105 
\citep{chen+01}. The latter authors derive an orbital period $P$ = 0.0584(2) d
from time-resolved spectroscopy. They furthermore find a strong orbital
modulation in a photometric light curve from 1991, and report a possibly
transient feature resembling a shallow eclipse.

We have observed the system in one night 5 years later, with the data set 
covering roughly 2.25 orbital cycles. The light curve is presented in Fig.\
\ref{j2315lc_fig}. The AOV periodogram yields two broad peaks corresponding to
$P$ and $P$/2. Since with our data set we cannot 
obtain the accuracy of \citet{chen+01} in the period 
determination, we have used their value to compute a phase-folded light curve
(Fig.\ \ref{j2315phi_fig}).  

The presence of strong flickering has a distorting effect within an individual
orbital cycle. However, in our average light curve we find the same principal
features as in Chen et al.'s data (their Fig.\ 3). Of special interest is that
we also detect an eclipse-like dip just after the primary maximum. Chen et al.\
report that this feature is not found in all their light curves. Also in our
data the first eclipse feature is much less evident than the second one.
By simply taking the average in time for the two data points that define
the second eclipse, and which have the same magnitude within the errors, we
obtain an ephemeris
\begin{equation}
T_0 = {\rm HJD}~2\,450\,363.711\,90(53) + 0.058\,4(2)~E,
\label{j2315eph_eq}
\end{equation}
with the periodic term taken from \citep{chen+01}. The error for the zero cycle
corresponds to half of the separation of the two data points, and is therefore 
likely to be somewhat overestimated.

\subsection{CTCV J2354$-$4700}

\begin{figure}
\rotatebox{270}{\resizebox{6cm}{!}{\includegraphics{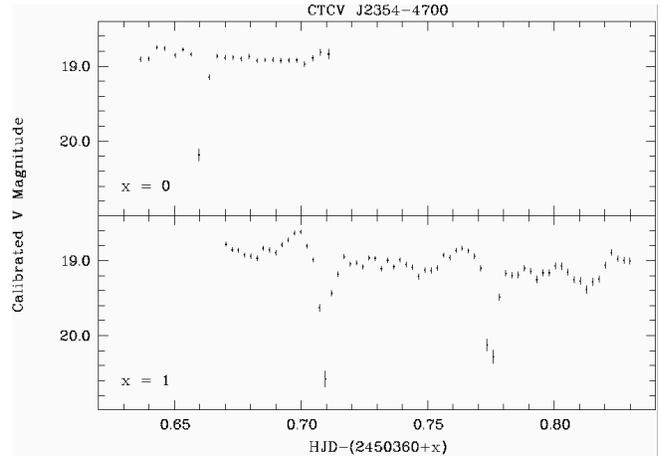}}}
\caption[]{Light curves for CTCV J2354$-$4700 from 1996-10-03 (top) and
1996-10-04 (bottom).} 
\label{j2354lc_fig}
\end{figure}

\begin{table}
\caption[]{Times of mid-eclipse for the CTCV J2354$-$4700 data.}
\label{j2354ecl_tab}
\begin{tabular}{l l}
\hline
\hline
cycle & HJD \\
\hline
0  & 2\,450\,360.661\,45(56) \\
16 & 2\,450\,361.709\,80(47) \\
17 & 2\,450\,361.775\,80(46) \\
\hline
\end{tabular}
\end{table}

\begin{figure}
\rotatebox{270}{\resizebox{6cm}{!}{\includegraphics{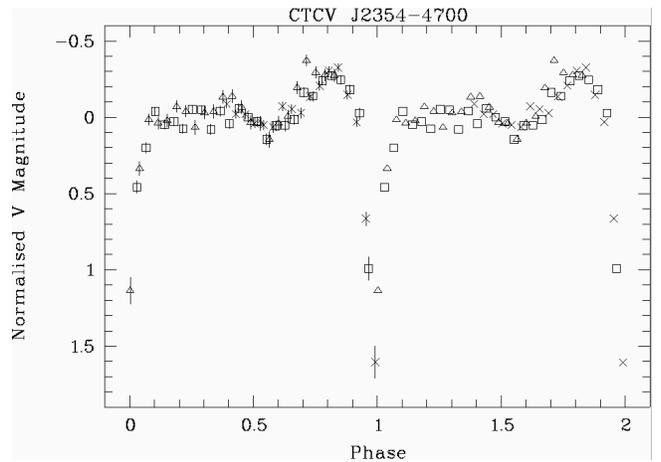}}}
\caption[]{The 1996-10-04 light curve of CTCV J2354$-$4700 folded on the
ephemeris given in Eq.\ \ref{j2354eph_eq}. The symbols in the plot refer to
different orbital cycles (15 = $\times$, 16 = $\Box$, 17 = $\triangle$).
The data have been normalised by subtracting a linear fit to the data 
(excluding humps and eclipses) with a slope of 2.59 mag d$^{-1}$.} 
\label{j2354phi_fig}
\end{figure}

The spectrum of this object is very similar to the one of CTCV J2315$-$3048,
i.e.\ of a dwarf nova in quiescence (Fig.\ \ref{sp_fig}).

CTCV J2354$-$4700 was observed on two subsequent nights. 
Fig.\ \ref{j2354lc_fig} presents the resulting light curves, and
Table \ref{j2354ecl_tab} lists the times of mid-eclipse which were obtained
by fitting a second order polynomial to the eclipse data.

A linear fit to these times yielded the ephemeris
\begin{equation}
T_0 = {\rm HJD}~2\,450\,360.661\,43(56) + 0.065\,538(39)~E.
\label{j2354eph_eq}
\end{equation}

Both data sets for CTCV J2354$-$4700 show steadily declining light curves. 
Linear fits, excluding the humps and eclipses, yielded slopes of 1.05(49) mag 
d$^{-1}$ and 2.59(24) mag d$^{-1}$ for the 1996-10-03 and the 1996-10-04 data, 
respectively. Since on both dates the average magnitudes are very similar, it
is clear that at least the value for the first night does not reflect the
system's behaviour over a longer period of time. Therefore, either the CV
has undergone some slow, non-orbital, variations between the two nights, or
the decline (perhaps in both nights) is artificial. For the 1996-10-04 data the
airmass has been continuously increasing (from 1.05 to 1.58), and 3 of the 5 
stars in the average comparison light curves are very red with $B-V \ge 0.9$, 
raising the suspicion that the decline may be due to differential extinction. 
However, light curves for the target computed with respect to the bluest 
($B-V = 0.4$) and the reddest ($B-V = 1.5$) comparison star yield nearly 
identical decline rates. Computing the difference light curve of the two 
comparison stars with the most extreme colour coefficient we do find the 
signature of differential extinction in the form of a decline of $\sim$0.1 mag 
d$^{-1}$ for both nights (in the first night, the standard deviation of the 
slope even still includes a constant light curve). In the average light curve, 
this effect can be assumed to be even less pronounced. We therefore do not
find any indication for an artificial origin of the observed decline, but due
to the rather limited amount of data we have to leave the question on the cause
of this apparent long-term variation to be answered by future studies.

In order to fold the data on the above ephemeris, we have subtracted the
linear decline from the 1996-10-04 data. Fig.\ \ref{j2354phi_fig} presents the 
such normalised, phase-folded, light curve. Its shape is very similar to that 
of CTCV J1300$-$3052, although in the latter both the amplitude of the hump and
the depth of the eclipse are roughly twice as strong, suggesting a lower 
inclination for CTCV J2354$-$4700.

\section{Summary}

\begin{table}
\caption[]{Summary of the orbital periods for the six CVs.}
\label{per_tab}
\begin{tabular}{l l l l}
\hline
object & $P$ [d] & $P$ [h] & notes \\
\hline
CTCV J0549$-$4921 & 0.080\,218(70) & 1.9252(17) & \\
CTCV J1300$-$3052 & 0.088\,981(32) & 2.135\,54(77) & [1]\\
CTCV J1928$-$5001 & 0.070\,179(01) & 1.684\,296(24) & [1] \\
CTCV J2005$-$2934 & 0.062\,887(37) & 1.5093(09) & [2] \\
CTCV J2315$-$3048 & 0.058\,4(2) & 1.402(05) & [1], [3] \\
CTCV J2354$-$4700 & 0.065\,538(39) & 1.5729(09) & [1] \\
\hline
\multicolumn{4}{l}{
\begin{minipage}{8cm}
[1] eclipsing, [2] most probable alias,
[3] period from \citet{chen+01}
\end{minipage}}\\
\end{tabular}
\end{table}

We have presented the results of time-series photometry for six CVs that were
detected in the Cal\'an-Tololo Survey, and where we find photometric variations
that we believe to be modulated with the orbital period. For five
of the systems we derived this previously unknown period, and in the 
case of CTCV J2315$-$3048 we have confirmed the existence of an eclipse, which
has been previously discussed as a possibly transient feature 
\citet{chen+01}. The resulting periods for all systems are summarised in
Table \ref{per_tab}. 


\section*{Acknowledgements}
This research is based on observations made at ESO, proposal numbers 57.D-0704,
58.D-0551, 58.D-0549, 59.D-0368. It has made use of the SIMBAD database, 
operated at CDS, Strasbourg, France. The Digitized Sky Surveys were produced at
the Space Telescope Science Institute under U.S. Government grant NAG W-2166, 
based on photographic data obtained using the Oschin Schmidt Telescope on 
Palomar Mountain and the UK Schmidt Telescope. IRAF is distributed by the 
National Optical Astronomy Observatories.
We thank the anonymous referee for helpful comments.

\appendix

\section{Finding charts}

We here present 5\arcmin$\times$5\arcmin~finding charts for the 6 CVs.

\begin{figure*}
\resizebox{14cm}{!}{\includegraphics{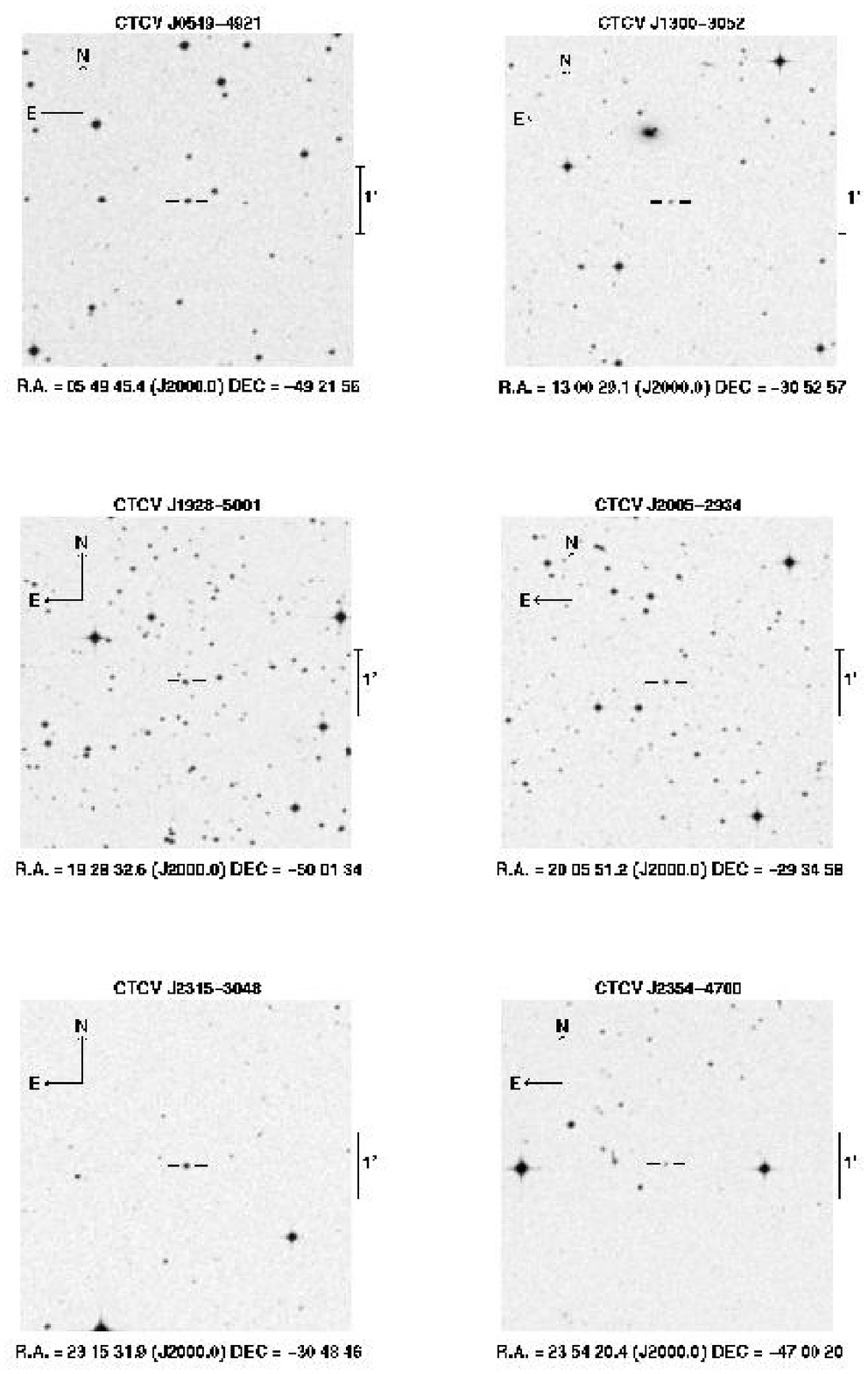}}
\caption[]{Finding charts from the Digitized Sky Survey.}
\label{fch_fig}
\end{figure*}

\end{document}